\documentclass[twocolumn,10pt]{IEEEtran}
\usepackage{amsmath}
\usepackage{subfigure}
\usepackage{epsf}
\usepackage{url}
\usepackage[usenames]{color}
\usepackage{graphicx}
\newtheorem{proposition}{Proposition}
\newtheorem{lemma}{Lemma}
\centerfigcaptionstrue

\begin{document}

\title{The Capacity of Random Ad hoc Networks under a Realistic Link Layer Model}
\author{Vivek P.~Mhatre\thanks{Vivek P.~Mhatre is with the School of Electrical and Computer Engineering,
Purdue University, West Lafayette, USA. Email: mhatre@ecn.purdue.edu}, Catherine P.~Rosenberg
\thanks{Catherine P.~Rosenberg is with the Department of Electrical and Computer Engineering, University of Waterloo, Canada.
Email: cath@ece.uwaterloo.ca}
}


\maketitle

\begin{abstract}
The problem of determining asymptotic bounds on the capacity of a random ad hoc network
is considered. 
Previous approaches assumed a threshold-based link layer model in which 
a packet transmission is successful if the SINR at the receiver is greater than a
fixed threshold.
In reality, the mapping from SINR to packet success probability is continuous.
Hence, {\em{over each hop}}, for every finite SINR, there is a non-zero probability of packet loss.
With this more realistic link model, it is shown that for a broad class of
routing and scheduling schemes, a {\em{fixed fraction of hops}} on each route   
have a {\em{fixed non-zero packet loss probability}}.
In a large network, a packet
travels an asymptotically large number of hops from source to destination. 
Consequently, it is shown that the cumulative effect of per-hop packet loss 
results in a per-node throughput of only $O\left(\frac{1}{n}\right)$
(instead of $\Theta\left(\frac{1}{\sqrt{n\log{n}}}\right)$ as shown previously for the threshold-based
link model).

A scheduling scheme is then proposed to counter this effect.
The proposed scheme improves the link SINR by using conservative spatial reuse, and improves the per-node
throughput to $O\left(\frac{1}{K_n\sqrt{n\log{n}}}\right)$, where each cell
gets a transmission opportunity at least once every $K_n$ slots, and $K_n\rightarrow\infty$ as
$n\rightarrow\infty$.
\end{abstract}

\section*{Keywords}
Ad hoc networks, capacity, SINR, interference, packet error, scheduling

\section{Introduction}
Recently, there has been considerable interest in the field of ad-hoc networks due to the vast
range of applications that they offer in several areas such as home networking, disaster recovery
networks, military, etc. The important question of the capacity of such ad-hoc networks was first
systematically analyzed by Gupta and Kumar in \cite{gupta}. This work has motivated several works 
under additional assumptions such as node mobility \cite{tse},
presence of limited infrastructure \cite{towsley}, capacity-delay
trade-offs \cite{kashyap,gaurav}, network information theoretic view of network
capacity \cite{liang}, etc.

In \cite{gupta}, Gupta and Kumar derive asymptotic
bounds on the capacity of a random ad hoc network.
In a random ad hoc network, nodes are deployed randomly and uniformly over the surface of a
sphere of unit area. Each node picks a
random node as its destination node, and sends packets to that node by using
multi-hop communication. All the nodes use a common transmit power level.
To derive their results, the authors assume that the region is covered with a tessellation of cells.
The cells are assumed to have certain uniformness properties in terms of size and shape. 
Such a uniform tessellation
simplifies the capacity analysis, since it is easier to account for routing and scheduling when the cells
have uniform size and shape.
The authors then demonstrate that under certain simplifying link layer assumptions for
successful packet reception, each node can achieve a throughput of $\Theta\left(
{\frac{1}{\sqrt{n\log{n}}}}  \right)$ packets
per second.
The authors also provide a scheduling and routing strategy in order to achieve this throughput.
The authors assume a link layer model in which, if the Signal to
Interference and Noise Ratio (SINR) at the receiver is greater than a certain threshold $\beta$, then
the packet is received successfully by the receiver with probability one.

In reality, for a given modulation and coding scheme, as long as there is some noise and interference, i.e., as long as
the SINR is finite, there is always a non-zero probability of packet error, and this probability of
error approaches zero as the SINR approaches infinity \cite{gallager,rappaport}. 
In other words, over {\em each hop}, the mapping from SINR to packet success probability is a continuous function that approaches one
only asymptotically as the SINR approaches infinity.
While the threshold-based packet reception model used in \cite{gupta} is a reasonable choice for successful packet reception in a single hop network such as a
cellular network, we argue that it needs to be refined when applied to a multi-hop network. In a
cellular network, if the SINR threshold $\beta$ is chosen to be sufficiently high, then the packet
error rate between the mobile and the base station (and vice versa) can be made very small
for a given choice of modulation and coding.
However, in the context of an ad hoc network, we know that each packet traverses 
{\em multiple hops}. The links of these hops receive interference from other ongoing transmissions
which could potentially corrupt the packet transmission over the  given link.
When a packet is relayed over a large number of links, each of which is likely to drop the packet
with a certain probability, the end-to-end throughput depends on the probability of end-to-end packet
delivery.
In \cite{gupta}, the capacity determining constraint is the 
number of source-destination paths that pass through a cell, i.e., the relaying burden of a cell.
However we believe that other factors such as interference, and the end-to-end packet error probability 
due to interference on each individual hop, should also be
taken into account when determining the network capacity.

Thus, our work is motivated by the fact that in a large ad-hoc network, a packet has to travel
over an asymptotically large number of links, each of which is unreliable with a certain probability.
We show that the existence of a $p<1$ that uniformly upper bounds the packet success probability of
at least a {\em fixed  fraction} of hops on each
source-destination path is sufficient to considerably degrade the network throughput. 
We then show the existence of such a $p<1$ for a broad range of routing and scheduling schemes
(including the one proposed in \cite{gupta}), and this results in an achievable per-node throughput
of just $O\left(\frac{1}{n}\right)$ instead of $\Theta\left({\frac{1}{\sqrt{n \log{n}}}}\right)$. 
Note that proving the existence of such a $p$ is not easy because, due to
the random node distribution it is possible that for certain hops, the transmitter and the receiver nodes
may be arbitrarily close to each other. This results in a very high SINR, and 
consequently a probability of packet success that is arbitrarily close to one.

To counter the throughput reduction due to this cumulative packet error effect,
we also show that by choosing a more conservative scheduling policy, we can 
reduce the interference (and thereby improve the packet success probability)
by trading off spatial reuse. This improves the per-node throughput to
$O\left(\frac{1}{K_n \sqrt{n \log{n}}}\right)$. Here $K_n$ is the length of the schedule, i.e., each cell
gets a transmission opportunity at least once every $K_n$ slots, and $K_n$ is chosen such that 
$K_n \rightarrow \infty$ as $n \rightarrow \infty$.

Since our arguments are mainly based on the impact of interference (via scheduling) and
packet error probability over long paths (asymptotically large number of hops), it is easy to see that the results are not restricted to a
specific communication paradigm such as the any-to-any communication model of \cite{gupta}. The {\em key
insight} is that if, in a network,
nodes use long multi-hop paths to reach their destination, and scheduling is used for exploiting spatial reuse, then
interference from other ongoing transmissions, and end-to-end error probability may degrade the
network throughput. Hence a trade-off between spatial reuse and cumulative packet error
probability has to be studied.

The rest of the paper is organized as follows. We motivate the problem, and outline our approach in
Section \ref{approach}. We discuss some of the related work in Section \ref{related}.
In Section \ref{prob}, we formulate and solve the capacity problem.
Finally, we present our conclusion and future directions in Section \ref{future}.

\section{Motivation and Approach}
\label{approach}
The arguments in this section are only meant to provide intuitive insights about the problem and our approach. 
We provide precisely proved results for all the arguments in Section \ref{prob}.
In \cite{gupta}, the authors study the problem of determining the capacity of random ad hoc networks.
They propose a routing and scheduling scheme to achieve a per-node throughput
of $\Theta\left({\frac{1}{\sqrt{n \log n}}}\right)$. 
It can be shown that for this scheme a packet traverses $\Theta\left( \sqrt{\frac{n}{\log n}}\right)$ intermediate hops
on its way from a source node to its destination node.
Thus asymptotically, the number of hops that a packet has to travel from source to destination goes
to infinity as $n$ scales.
The authors assume that if the SINR over a hop is at least $\beta$,
the packet transmission is successful. Let us call this the ideal link model.
The authors then use a scheduling scheme which ensures an SINR of at least $\beta$ over all the
hops. Thus, under the ideal link model, and the proposed scheduling policy, {\em every}
packet transmission is successful.

Now consider another link model in which a link is reliable with a probability $p<1$, i.e.,
each packet transmitted on the link is received successfully at the receiver with a probability of $p$.
Let us call this the probabilistic lossy link model.
For simplicity, assume that over each hop, if a packet is not received successfully
by the receiver, the transmitter does not retransmit the packet, and the packet is lost.
Also assume that we use the same routing and scheduling policy as used above for the ideal link model.
Since there are  $\Theta\left( \sqrt{\frac{n}{\log n}}\right)$ hops from source to destination,
the probability that the packet reaches its destination scales as $p^{\sqrt{\frac{n}{\log n}}}$.
It is easy to show that this quantity is $O\left(\sqrt{\frac{\log n}{n}}\right)$, since $p^m$ is
$O\left(\frac{1}{m}\right)$ as $m$ tends to infinity.
Let us consider two identical networks, one with the ideal link model, and the other with the 
probabilistic lossy link model, and assume that for both of them, we use the same routing and scheduling
algorithm as outlined in \cite{gupta}. The only difference between the two networks is that while in
the former case, all the packets injected by the source into the network reach the destination, in
the latter case, only a fraction of the injected packets reach the destination.
Hence, under the probabilistic lossy link model, the achievable throughput is the product of the achievable
throughput of the ideal link model, $\Theta\left({\frac{1}{\sqrt{n \log n}}}\right)$, and
an end-to-end packet delivery probability term that is $O\left(\sqrt{\frac{\log n}{n}}\right)$.
This results in an end-to-end throughput of $O\left(\frac{1}{n}\right)$ for the probabilistic lossy
link model.

A realistic link model lies somewhere between the ideal link model and the probabilistic lossy
link model. In a realistic link model, the probability of link reliability, $p$, is a continuous
function of the SINR
of the link. The SINR in turn depends on factors such as transmitter-receiver separation, power of
interference from simultaneous transmissions, and noise power. These factors are different for
different links along a path. 
For example, since the nodes are randomly placed, it is possible that for a given hop, the two
nodes could be arbitrarily close to each other resulting in a very high SINR for that link.
This in turn means that the packet success probability for that link could be arbitrarily close
to one. 
Thus, there is no single $p <1$ which uniformly upper bounds the probability of
packet success for all the links.
However, if for at least a fixed fraction of links along every source-destination path,
there is a $p<1$ which upper bounds the probability of successful packet reception,
then the per-node throughput bound of $O\left(\frac{1}{n}\right)$ which holds for
the probabilistic link model also holds for the realistic link model.
To get around this problem we must design scheduling (and/or retransmission
strategies) so that no such $p$ exists.

The above arguments are made more precise in Section \ref{prob}.

\section{Related Work}
\label{related}
Throughout this paper, we refer to the work of Gupta and Kumar on the
capacity of random ad hoc networks \cite{gupta}. In this work, the
authors assume a simplified link layer model in which each packet
reception is successful if the receiver has an SINR of at least $\beta$.
The authors assume that each packet is decoded at every hop along the path from source to
destination. No co-operative communication strategy is used, and
interference signal from other simultaneous transmissions is treated
just like noise. For this communication model, the authors propose a
routing and scheduling strategy, and show that a per-node throughput of
$\Theta \left( \frac{1}{\sqrt{n \log{n}}}\right)$ can be achieved.

In \cite{liang}, the authors discuss the limitations of the work in
\cite{gupta}, by taking a network information theoretic approach. The
authors discuss how several co-operative strategies such
as interference cancellation, network coding, etc. could be used to
improve the throughput. However these tools cannot be exploited fully
with the current technology which relies on point-to-point coding, and
treats all forms of interference as noise. The authors also discuss how
the problem of determining the network capacity from an information
theoretic view-point is a difficult problem, since even the capacity of
a three node relay network is unknown. In Theorem 3.6 in \cite{liang},
the authors determine the same bound on the capacity of a random network as 
obtained in \cite{gupta}.

In \cite{gupta}, the authors consider another model of ad hoc networks called arbitrary networks.
In an arbitrary network, nodes are placed at arbitrary
locations in a region of fixed area, source-destination pairs are chosen arbitrarily, and each node can
transmit at an arbitrary power level. This is unlike a random ad hoc network where node are deployed
randomly and uniformly, source-destination pairs are chosen at random, and all the nodes transmit at
the same power level\footnote{The definitions of random networks and arbitrary networks have been taken from 
\cite{gupta}.}. For the arbitrary network model, the authors show that a per-node throughput of
$\Theta\left(\frac{1}{\sqrt{n}}\right)$ is achievable. 
However, for random networks, the routing and scheduling scheme proposed in
\cite{gupta} only achieves a throughput of $\Theta\left(\frac{1}{\sqrt{n\log{n}}}\right)$. This gap
between the arbitrary networks and the random networks is closed in \cite{franc}. In \cite{franc}, the
authors note that the requirement of connectivity with high probability in random networks as required
in \cite{gupta}, requires higher transmission power at all the nodes, and results in excessive
interference. This in turn lowers the throughput of random networks from $\Theta\left(\frac{1}{\sqrt{n}}\right)$ to
$\Theta\left(\frac{1}{\sqrt{n\log{n}}}\right)$. Motivated by this, the authors in \cite{franc}
propose using a backbone-based relaying scheme in which instead of ensuring connectivity with
probability one, they allow for a small fraction of nodes to be disconnected from the backbone. 
The nodes in the backbone are densely connected, and can communicate over each hop at a constant rate. 
Such a backbone traverses up to $\sqrt{n}$ hops. The nodes that are not a part of the backbone, send
their packets to the backbone using single hop communication. The authors then show that the
interference caused by these long range transmissions does not impact the traffic carrying capacity of
the backbone nodes. Finally, the authors show that the relaying load of the backbone determines 
the per-node throughput of such a scheme, and this results in an achievable per-node throughput of
$\Theta\left(\frac{1}{\sqrt{n}}\right)$. The above approach of allowing a few disconnected nodes in the network has also been used in
\cite{dousse}.

However, just as in \cite{gupta},  all the above mentioned works assume that
over each link a certain non-zero rate can
be achieved. They do not take into account the fact that in reality, 
such a rate is achieved {\em with a probability of bit error arbitrarily
close (but not equal) to one}. Once the coding and
modulation scheme is fixed, the function corresponding to the
probability of bit error is also fixed.

\section{Problem Formulation and Solution}
\label{prob}
We use the same terminology as used by Gupta and Kumar in \cite{gupta}. 

\begin{itemize}
\item $f_n = \Theta\left(g_n\right)$, if there exist constants $a_1$ and $a_2$ such that, for $n$ large enough,
$a_1 g_n \leq f_n \leq a_2 g_n$
\item $f_n = \Omega\left(g_n\right)$, if there exists a constant $a_1$ such that, for $n$ large enough,
$a_1 g_n \leq f_n $
\item $f_n = O\left(g_n\right)$, if there exists a constant $a_2$ such that, for $n$ large enough,
$f_n \leq a_2 g_n$
\item $f_n = \omega\left(g_n\right)$, if there exists a constant $a_1$ such that, for $n$ large enough,
$a_1 g_n < f_n$
\item $f_n = o\left(g_n\right)$, if there exists a constant $a_2$ such that, for $n$ large enough,
$f_n < a_2 g_n$
\end{itemize}
Consider
a sphere of unit area, say $S^2$, over which $n$ nodes are deployed randomly and uniformly. Each node picks a
random node which is its destination node, and sends packets to this destination node. Each node has a
common transmit power level $P$, and it uses intermediate nodes as relays to reach the destination node. 
The surface of the sphere is covered by a Voronoi tessellation in such a way that each Voronoi
cell $V$ can be enclosed inside a circle of radius $2\rho_n$, and each circle encloses a circle of
radius $\rho_n$ (all the distances are measured along the surface of $S^2$). We use such a tessellation
to ensure uniform cell size. 
The cells can be made arbitrarily small in a uniform way by choosing $\rho_n$ small. In \cite{guptaconn}, Gupta
and Kumar have derived necessary and sufficient conditions for asymptotic connectivity of a random ad-hoc network.
By using results from \cite{guptaconn} in \cite{gupta}, the authors choose $\rho_n$ to be the radius of a circle
of area $\frac{100\log{n}}{n}$ on $S^2$. With this choice of $\rho_n$, and with the communication
radius of each node chosen to be $8\rho_n$, the authors show that all the nodes in the network are
connected with probability approaching one as $n$ approaches infinity. 
 
The authors in \cite{gupta}
use the following criterion for successful packet reception. If 
the SINR at a receiver is greater than a certain fixed threshold
$\beta$, then the packet is successfully decoded by the receiver. In other words, a transmission from
node $i$ to node $j$ is successful if:
\begin{equation}
\label{model}
\frac{ \frac{P}{{|X_i - X_j|}^{\alpha}}}{ N + \sum\limits_{k\in T \mbox{, } k\neq i}\frac{P}{{|X_k - X_j|}^{\alpha}}}
\geq \beta \mbox{, }
\end{equation}
where $T$ is the set of all the nodes that are transmitting simultaneously with node $i$, and $\alpha
> 2$ is the propagation loss exponent.

In our analysis, to begin with we make the following assumptions:
\begin{description}
	\item[A1] The region $S^2$ is partitioned into a Voronoi tessellation such that each cell contains a circle of radius
	$\rho_n$, and each cell is contained inside a circle of radius $2\rho_n$.
	We assume that $\rho_n$ is at least $\Omega\left(\sqrt{\frac{\log{n}}{n}}\right)$ to ensure network	connectivity with high probability.
  To ensure that the cell sizes shrink as $n$ scales, $\rho_n \rightarrow 0$ as $n$ approaches infinity. 
	Each node transmits at a fixed power level $P$.
	\item[A2] For a given modulation and coding scheme, 
	the probability of a successful packet reception is a continuous increasing  function
  of SINR that approaches one as the SINR approaches infinity. Hence, for every finite SINR value, we
  assume that there is a non-zero probability of packet loss. Even as $n$ scales, the modulation and
	coding scheme is fixed. We also assume that over each hop, if a packet is not received successfully
by the receiver, the transmitter does not retransmit the packet, and the packet is lost.
	\item[A3] A scheduling algorithm that guarantees each cell
	a transmission opportunity at least once every $K$ slots, where $K$ is a finite number that is
	independent of $n$. In other words, the length of the schedule is bounded even as $n$ scales.
	In \cite{gupta}, it was shown that such a scheduling strategy guarantees an SINR of at least
	$\beta$ for all the scheduled transmissions.
	\item[A4] A routing scheme in which packets are routed along straight line paths between 
	source-destination pairs, i.e., every cell that intersects the straight
	line joining a source-destination pair, relays the packets of that pair.
\end{description}

The key assumption that distinguishes our work from the previous approaches is assumption A2.
The remaining three assumptions are also used by Gupta and Kumar in \cite{gupta}.
We later on relax assumptions A3 and A4, and also comment on relaxing assumption A2.
Note that even with A3 and A4, the assumptions above are general enough to encompass a vast array of routing and scheduling policies
that could be used for multi-hop communication. 

\subsection{{Per-node Capacity under Assumptions A1 to A4}}
\label{seca1a5}
Consider Fig.~\ref{figH}. Let $L_i$ be the line segment along the surface of $S^2$ that connects the $i$th
source-destination pair (henceforth referred to as the $i$th connection).
We also use $L_i$ to denote the length of the line segment
joining the $i$th source-destination pair. As per A4, the packets of the $i$th
connection are relayed hop-by-hop by every cell which intersects line $L_i$. Over each hop, any node
in the relaying cell may forward the packet. The scheduling algorithm, and the uniform cell sizes
ensure that communication between any two nodes in the neighboring cells is possible by guaranteeing
that the SINR at the receiving node is greater than or equal to $\beta$ (see \cite{gupta} for more
details). 

We will now prove that for each source-destination pair, there are at least  $\Theta\left(
{\frac{1}{\rho_n}}\right)$ hops over which the SINR is
upper bounded by a fixed constant.
For this, we need the next three lemmas.

\begin{lemma}
\label{lemmaH}
For the routing scheme in A4, the number of hops $H_i$ for connection $i$ is
$\Theta\left( \frac{L_i}{\rho_n} \right)$. More precisely,
\begin{equation*}
\frac{1}{8} \frac{L_i}{\rho_n} \leq H_i \leq \frac{16}{\pi} \frac{L_i}{\rho_n}\mbox{.}
\end{equation*}
\end{lemma}
\vspace{0.2cm}

\begin{proof}
Since, each Voronoi cell is contained in a circle of radius $2 \rho_n$, the maximum distance between
a point in a given cell, and a point in its neighboring cell is $8 \rho_n$ (see Fig.~\ref{figH}). 
Thus, over each hop, the
maximum distance that a packet can cover is upper bounded by $8 \rho_n$. Hence the lower bound.
\begin{figure}
\epsfysize=3.25cm
\centerline{\epsfbox{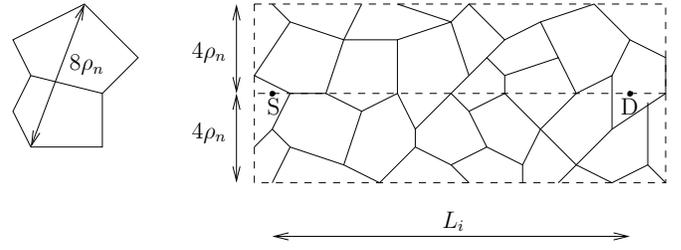}}
\caption{Bounding the number of hops $H_i$ for connection $i$.}
\label{figH}
\end{figure}

For the upper bound, if we look at a strip that is $4 \rho_n$ wide on either sides of $L_i$ 
(see Fig.~\ref{figH}), we observe that if a
cell is used as a relay cell, then it must lie {\em entirely}\, within this strip. This is because
each cell is completely contained inside a circle of diameter $4 \rho_n$. The area of this strip is
$8 \rho_n L_i$. Each cell contains a circle of radius $\rho_n$, and hence the area of each cell is at
least $\pi {\rho_n}^2 / 2$ (see Lemma \ref{lemmaEXP2} in Appendix). Thus, the maximum number of
cells that lie {\em entirely}\, within this strip
is upper bounded by $16 L_i / \pi \rho_n$. Hence the upper bound.

The lower bound is actually $\max \left( \frac{L_i}{8 \rho_n}, 1\right)$, but with
$L_i=\Theta(1)$, $\rho_n$ is small for $n$ large, and we can ignore one. 
In any case, we require the upper bound, and not so much the lower bound in our subsequent analysis.
Also note, that for a more exact upper bound, we should consider a strip of length $L_i + 4 \rho_n + 4
\rho_n$ instead
of $L_i$, since the source and destination nodes could be located at the peripheries of their
respective cells. However, since $L_i$ is $\Theta (1)$, and $\rho_n$ approaches zero for $n$ large,
we can safely ignore the two terms corresponding to the sizes of the source and destination cells.
\end{proof}
\vspace{0.2cm}

\begin{lemma}
\label{lemmaHh}
Fix $t$ such that $0 < t < 1$. For connection $i$, out of $H_i$ total hops, let $h_i$ hops be such that each of
these hops covers a distance of less than $t \rho_n$. Then
\begin{equation*}
H_i - h_i \geq \frac{L_i}{\rho_n} \left(\frac{1 - \frac{16 t}{\pi}}{8 - t} \right)
\end{equation*}
Thus, for the above $H_i - h_i$ hops, the signal received at the receiver is at the most $P (t
\rho_n)^{-\alpha}$, where $P$ is the transmit power common to all the nodes, and $\alpha$ is the propagation loss exponent ($\alpha > 2$).
\end{lemma}
\vspace{0.2cm}

\begin{proof}
Since $h_i$ hops each cover less than $t \rho_n$ distance, the leftover distance, which is at least
$L_i - h_i t \rho_n$ has to be covered by the rest of the $H_i - h_i$ hops. Each of these $H_i - h_i$
hops can cover a distance of at the most $8 \rho_n$. Hence, we have
\begin{align*}
(H_i - h_i)\, 8 \rho_n &\geq L_i - h_i(t \rho_n) \\
\Rightarrow h_i &\leq H_i \left( \frac{8}{8-t} \right) - \frac{L_i}{\rho_n}
\left(\frac{1}{8-t}\right) \\
\Rightarrow H_i - h_i &\geq H_i \left(\frac{-t}{8-t}\right) + \frac{L_i}{\rho_n} \left(\frac{1}{8-t}\right)\\
 &\geq \frac{L_i}{\rho_n} \left(\frac{16}{\pi}\right) \left(\frac{-t}{8-t}\right) +
 \frac{L_i}{\rho_n}\left(\frac{1}{8-t}\right)\mbox{,}
\end{align*}
where we have used Lemma \ref{lemmaH} to upper bound $H_i$ on the right hand side in the last step.
Thus,
\begin{equation*}
H_i - h_i \geq \frac{L_i}{\rho_n}\left(\frac{1 - \frac{16 t}{\pi}}{8-t}\right) 
\end{equation*}
\end{proof}
\vspace{0.2cm}

In \cite{gupta}, it was shown that there
exists a scheduling policy that ensures an SINR of at least $\beta$ at the receiver of
every scheduled transmission. 
This scheduling policy corresponds to a graph coloring problem, and it was
shown in \cite{gupta} that the maximum number of colors required to color all the cells 
is upper bounded by $1 + c_1$, where $c_1$ is a fixed constant that is independent of $n$ (see Lemma 4.4 in \cite{gupta}).
Using this scheduling scheme, each cell gets a transmission opportunity at least once every $1 + c_1$
time slots. Thus, with respect to A3, $K=1+c_1$.

The objective of the following lemma is to show that except for a small fraction of hops, all the
remaining hops of a connection receive a certain minimum amount of interference from other
simultaneous transmissions.
\begin{lemma}
\label{lemmaM}
Fix $M>9$.
Let $N_i$ be the number of hops of connection $i$ such that
there is no simultaneous interfering transmission within a circle of radius $(M+8) \rho_n$ around
the receivers of those hops. To avoid tedious boundary conditions, let us not count the hops containing the source
and the destination nodes in $N_i$. Then,
\begin{equation*}
N_i \leq \frac{L_i}{\rho_n} \left( \frac{2(1 + c_1)}{M}\right)\mbox{,}
\end{equation*}
where $c_1$ is the constant from Lemma 4.4 in \cite{gupta}.
\end{lemma}
\vspace{0.2cm}

\begin{proof}
There are at the most $1+c_1$ colors, i.e., each cell
gets a transmission opportunity at least
once every $1+c_1$ time slots. Let us index these $1+c_1$ colors, and consider any one of these colors,
say color $j$. 
We would like to determine the number of color-$j$ cells of connection $i$ that are at least 
$M \rho_n$ distance away from other color-$j$ cells.

For this, let $U_i^j$ be the set of all the cells of color $j$ that a packet of connection $i$ traverses.
We refer to a circle as a ``surrounding circle of a cell'' if its {\em center lies within the cell}.
Consider all the surrounding
circles of radius $M \rho_n$ around each of the cells in set $U_i^j$. 
For each cell, there are many such circles, since the only requirement is that
the center of the circle lie within the cell. A subset $V_i^j$ of $U_i^j$ is then formed as follows.
A cell from set $U_i^j$ is added to set $V_i^j$ if {\em none} of
its surrounding circles of radius $M \rho_n$ contains  any other color-$j$ cell either partially or fully.
Thus every cell in set $V_i^j$ is at least $M \rho_n$ distance away from any other color-$j$ cell.

\begin{figure}
\epsfysize=3cm
\centerline{\epsfbox{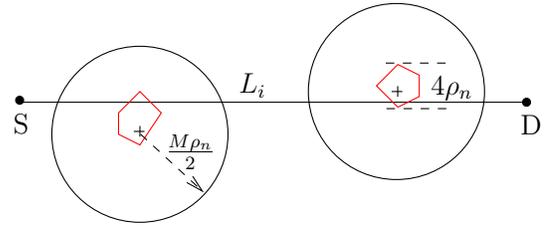}}
\caption{Bounding the number of hops $N_i$ of connection $i$ for which there is no interferer within
a distance $(M+8) \rho_n$ from the receiver.}
\label{figM}
\end{figure}

Hence, any surrounding circle of radius $\frac{M \rho_n}{2}$ around a cell in set $V_i^j$ does not 
overlap with any surrounding circle of radius $\frac{M \rho_n}{2}$ around another cell in set $V_i^j$.
Let there be $N_{i}^{j}$ cells in set $V_i^j$. 
The total length of the path is $L_i$ (referring to Fig.~\ref{figM}). We would like to determine the
maximum number of {\em non-overlapping surrounding}\, circles of radius $\frac{M \rho_n}{2}$ around color-$j$ cells that can be
accommodated along $L_i$. This number is an upper bound for $N_i^j$.
Each cell along $L_i$ can be enclosed inside a circle of diameter $4
\rho_n$. Referring to Fig.~\ref{figM}, maximal packing of the non-overlapping circles of radius $\frac{M \rho_n}{2}$ occurs when the
chord formed by the intersection of $L_i$ with these circles is of minimal length. This occurs when the
chord is shifted from its diametrical position by $4 \rho_n$. This is the maximum distance that the
chord can be shifted while still ensuring that $L_i$ intersects the cell, and the center of the
circle of radius $\frac{M \rho_n}{2}$ surrounding the cell lies inside the cell. The
length of each chord formed by the intersection of $L_i$ and the surrounding circle of radius $\frac{M \rho_n}{2}$ is at
least $2\left( \sqrt{\left(\frac{M \rho_n}{2}\right)^2 - \left(4 \rho_n\right)^2} \right)$,
which in turn is at least $\frac{M \rho_n}{2}$ if $M> 9$. Since $N_{i}^{j}$ such chords are to 
be accommodated along the length of the path,
\begin{equation*}
N_{i}^{j} \frac{M \rho_n}{2} \leq L_i \Rightarrow N_{i}^{j}  \leq \frac{2 L_i}{M\rho_n}
\end{equation*}
Therefore for all the colors,
\begin{align*}
N_i &= \sum\limits_{j=1}^{1+c_1} N_{i}^{j} \\
&\leq\frac{L_i}{\rho_n}\left( \frac{2(1+c_1)}{M} \right)
\end{align*}
Thus except for these $N_i$ hops, each of the hops of connection $i$ has at least one cell that has a
simultaneous ongoing transmission
within a radius of $M \rho_n$ of the given cell. Or equivalently, within a radius of $(M+4+4)
\rho_n = (M+8)\rho_n$ of the {\em receivers}\, of these hops, there is at least one more
node that is transmitting simultaneously. This is because the maximum size of a cell is $4 \rho_n$.
And this proves the result.
\end{proof}
\vspace{0.2cm}

Let $A_i$ be the set of hops of connection $i$ for which the received signal is at the most $P
(t \rho_n)^{-\alpha}$. Then, given $\epsilon_1 > 0$, we can find $t>0$ small enough so that using
Lemma \ref{lemmaHh}, 
\begin{equation}
\label{nAi}
|A_i| \geq \frac{L_i}{\rho_n} \left( \frac{1}{8} - \epsilon_1 \right)
\end{equation}
Let $B_i$ be the set of hops of connection $i$ for which there is no simultaneous transmissions within a distance of 
$(M+8) \rho_n$ of the receiver.
Using Lemma \ref{lemmaM}, $|B_i| = N_i$, and given $\epsilon_2 >0$, 
we can choose $M$ large enough so that
\begin{align}
|B_i| &\leq \frac{L_i}{\rho_n} \left( \frac{2(1+c_1)}{M}\right) \nonumber \\
      &\leq \frac{L_i}{\rho_n} \epsilon_2
\label{nBi}			
\end{align}
Thus using \eqref{nAi} and \eqref{nBi},
\begin{align*}
|A_i \cap {B_i}^{c}| &\geq |A_i| - |B_i| \nonumber \\
& \geq \frac{L_i}{\rho_n} \left( \frac{1}{8} - \epsilon_1\right) - \frac{L_i}{\rho_n} \epsilon_2 \nonumber \\
& = \frac{L_i}{\rho_n} \left( \frac{1}{8} - \epsilon_1 - \epsilon_2\right)
\end{align*}
If we pick $\epsilon_1 = \epsilon_2 = 1/32$, and choose $t=t_0$ and $M=M_0$ corresponding to this choice of
$\epsilon_1$, $\epsilon_2$, then 
\begin{equation}
|A_i \cap {B_i}^{c}| \geq \frac{L_i}{16 \rho_n}
\label{nAB}
\end{equation}
Note that $A_i \cap {B_i}^{c}$ is the set of hops over which the received signal is no more than $P
(t_0 \rho_n)^{-\alpha}$, {\it and} there is at least one simultaneous transmission within a distance of $(M_0
+ 8) \rho_n$ of the receiver. This in turn means that for these hops, the SINR is upper bounded by
\begin{align}
SINR &\leq \frac{P (t_0 \rho_n)^{-\alpha}}{P \left((M_0 + 8)\rho_n\right)^{-\alpha}} \nonumber \\
\label{sinr}
&= \left( \frac{M_0 + 8}{t_0} \right)^{\alpha} \\
&= \beta_0 
\end{align}
Thus we have proved the following Proposition.
\begin{proposition}
\label{propphi}
There exist fixed constants $t_0$ and $M_0$, that do not depend on $n$, such that for at least
$L_i/16\rho_n$ hops of connection $i$, the SINR is less than a fixed constant $\beta_0$ given by
\begin{equation*}
\beta_0    = \left( \frac{M_0 + 8}{t_0} \right)^{\alpha}
\end{equation*}
As per A2, since the SINR is upper bounded by a fixed constant $\beta_0$, the
probability of successful packet reception is also upper bounded by a fixed
constant $\phi(\beta_0) < 1$.
\end{proposition}
\endproof

Let us assume that time is slotted such that each slot is long enough to transmit a single packet of
fixed length. Let $\lambda_n (i)$ be the rate in packets/slot at which source node $i$ injects packets in the network.
Although the source node injects packets at a rate of $\lambda_n (i)$ packets per slot, not all the
packets make it to the destination node. This is because, at each hop, the SINR is finite, and hence
there is a non-zero probability of the packet getting dropped. The actual end-to-end throughput of
connection $i$, denoted by $\Lambda_n (i)$ is given by,
\begin{align*}
\Lambda_n (i) = \lambda_n (i)\, &\mbox{Prob} \left\{\mbox{packet is received successfully over}\right. \\
                              &\left.\mbox{all the hops of connection $i$}\right\} \nonumber \\
 = \lambda_n (i)\, &\Pi_{j=1}^{H_i} \mbox{Prob} \left\{\mbox{packet is received successfully}\right. \\
                 &\left.\mbox{over the $j$th hop of connection $i$}\right\}\mbox{,}
\end{align*}
where we assume that the interference signal, and noise observed by a packet at each of the hops
are independent. We know from Proposition \ref{propphi} that, among the $H_i$ hops of connection $i$, at least $L_i/16 \rho_n$ hops have a probability of packet
success of no more than $\phi(\beta_0)$. Thus,
\begin{equation}
\Lambda_n (i) \leq \lambda_n (i)\, \left\{\phi(\beta_0)\right\}^{\frac{L_i}{16\rho_n}}
\label{Lambdafirst}
\end{equation}
Note that $L_i$ are i.i.d. random variables. Hence if we remove the conditioning on $L_i$ by taking
the expectation with respect to $L_i$, the end-to-end throughput $\Lambda_n$ is,
\begin{align}
\Lambda_n &= {\bf E}_L [\Lambda_n (i)]  \\
          &\leq \lambda_n (i)\, {\bf E}_L \left\{
					\left(\left\{\phi(\beta_0)\right\}^{\frac{1}{16\rho_n}}\right)^{L_i} \right\}	\\
          \label{Lambda}
          &= \lambda_n (i)\, {\bf E}_L \left[ \delta^{L_i} \right]
\end{align}
where we have substituted $\delta = \left\{\phi(\beta_0)\right\}^{\frac{1}{16\rho_n}}$. 
Note that in determining the average end-to-end throughput $\Lambda_n$, we take expectations at two
levels; once to take into account the randomness due to the possibility of packet error on each link, and
once to take into account the randomness due to the locations of the source and destination nodes.
Also note that $0 < \delta < 1$. Since $L_i$ is a line connecting two points
picked at random on the surface of $S^2$, we can show that (See Lemma \ref{lemmaEXP} in Appendix).
\begin{equation}
\label{deltaL}
{\bf E}_L \left[ \delta^{L_i} \right] = \frac{2\pi\left(1 +
\delta^{\frac{\sqrt{\pi}}{2}}\right)}{4\pi + (\log \delta)^2}
\end{equation}
Using \eqref{Lambda} and \eqref{deltaL},
\begin{align}
\Lambda_n &\leq \lambda_n (i)\,  \frac{2\pi\left(1 + \delta^{\frac{\sqrt{\pi}}{2}}\right)}{(4\pi + (\log \delta)^2)} \nonumber \\
          &=   \lambda_n (i)\,   \frac{2\pi\left(1 +
					\left\{\phi(\beta_0)\right\}^{\frac{\sqrt{\pi}}{32\rho_n}}\right)}{\left(4\pi
					+ \left(\frac{1}{16\rho_n}
					\log {\phi(\beta_0)}\right)^2\right)} \nonumber \\
          &=  \lambda_n (i)\,    \frac{512\pi{\rho_n}^2\left(1 +
					\left\{\phi(\beta_0)\right\}^{\frac{\sqrt{\pi}}{32\rho_n}}\right)}{\left(1024\pi{\rho_n}^2
					+ (\log {\phi(\beta_0)})^2\right)} \nonumber \\
          &< \lambda_n (i)\,  \frac{1024\pi{\rho_n}^2 }{\left( 1024\pi{\rho_n}^{2} + 
					(\log {\phi(\beta_0)})^2\right)}\mbox{, } \quad \mbox{ since $\phi(\beta_0) < 1$.} \nonumber \\
          &< \lambda_n (i)\,  \frac{1024\pi{\rho_n}^2 }{(\log
					{\phi(\beta_0)})^2}
					\label{Lambdainter}
\end{align}
So far, i.e., in proving Lemmas \ref{lemmaH}, \ref{lemmaHh} and \ref{lemmaM}, Proposition
\ref{propphi}, and \eqref{Lambdainter} we have not made any assumptions about the exact form of $\rho_n$. We have just assumed A1 to A4.
Let us now study the per-node throughput bound for 
for the choice of $\rho_n$ in \cite{gupta}, i.e., when $\rho_n$ is chosen to be the radius of a disk
of area $\frac{100 \log{n}}{n}$.
In \cite{gupta}, it was shown through Lemma 4.8, that for this choice of $\rho_n$ each Voronoi cell contains at
least one node with
high probability. More precisely, there exists a sequence $\delta_n \rightarrow 0$, such that
\begin{align*}
\mbox{Prob} &\left\{\mbox{Number of nodes in cell $V$ $\geq$ $50\log n$, for every}\right.\\
&\left.\mbox{cell $V$ in the tessellation}\right\} > 1 - \delta_n
\end{align*}
Thus we have the following Lemma for the above choice of $\rho_n$.
\begin{lemma}
\label{lemma50logn}
If $\lambda_n (i)$ is the rate in packets/slot at which {\em every}\, source node injects packets in the network, then with high
probability,
\begin{equation*}
\lambda_n (i) \leq \frac{1}{50 \log n}
\end{equation*}
\end{lemma}
\vspace{0.2cm}

\begin{proof}
The scheduling algorithm proposed in \cite{gupta} guarantees each cell a time slot at least once every $1+ c_1$ slots. However,
since each cell contains at least $50 \log n$ nodes with high probability, even if each node were to
transmit only its own packets, and not relay packets, it would still get no more than one
transmission opportunity every $50 \log n$ slots.
Hence, the rate at which a node injects its own packets into the network can never be more than
$1/50\log{n}$.
\end{proof}
\vspace{0.2cm}

Using the above Lemma to bound $\lambda_n (i)$ in \eqref{Lambdainter} we get
\begin{equation}
\Lambda_n          <  \frac{1}{50 \log n} \frac{1024\pi{\rho_n}^2 }{(\log
					{\phi(\beta_0)})^2}
\label{Lambdalast}
\end{equation}
Also, since $\rho_n$ is the radius of a circle of area $\frac{100 \log{n}}{n}$, using Lemma \ref{lemmaEXP2} in Appendix we have,
\begin{align*}
\frac{\pi {\rho_n}^2}{2} &< \frac{100\log n}{n}
\end{align*}
Substituting the above in \eqref{Lambdalast},
\begin{equation}
\Lambda_n 
          = \frac{2^{12}}{(\log {\phi(\beta_0)})^2} \frac{1}{n}
					\label{final}
\end{equation}
Thus we have proved the following important Proposition.
\begin{proposition}
\label{propmain1}
Under assumptions A1 to A4, and with $\rho_n$ chosen to be the
radius of a disk of area $\frac{100 \log{n}}{n}$ (as in \cite{gupta}), the per-node throughput that can be achieved, $\Lambda_n$,
is $O(\frac{1}{n})$ instead of $\Theta\left(\frac{1}{\sqrt{n \log{n}}}\right)$. More precisely,
\begin{equation*}
\Lambda_n \leq \frac{c_0}{n} \mbox{, }
\end{equation*}
where $c_0 = \frac{2^{12}}{(\log {\phi(\beta_0)})^2}$ is a constant that does not depend on $n$.
\end{proposition}
\endproof

Note that for keeping all the calculations simple, we have tried to ensure that the constant in the
asymptotic expression is a power of two. However, a tighter bound with a smaller asymptotic constant
can be easily obtained.

The upper bound of $O\left(\frac{1}{n}\right)$ on per-node throughput
can also be extended to any other choice of $\rho_n$ that satisfies assumption A1.
\begin{lemma}
Consider a tessellation that satisfies A1 with a certain $\rho_n$.
Then there exists a sequence $\delta_n \rightarrow 0$, such that
\begin{align*}
\mbox{Prob} &\left\{\mbox{Number of nodes in cell $V$ $\geq$ $\frac{\pi n {\rho_n}^2}{4}$, for every}\right.\\
&\left.\mbox{cell $V$ in the tessellation}\right\} > 1 - \delta_n\mbox{, }
\end{align*}
i.e., with high probability each cell contains at least $\frac{\pi n {\rho_n}^2}{4}$ nodes.
\end{lemma}
\vspace{0.2cm}
\begin{proof}
Let $a_n$ be the area of a disk of radius $\rho_n$.  We
can use similar arguments as in Lemma
4.8 in \cite{gupta}. However instead of inscribed disks of area $\frac{100 \log(n)}{n}$, we have
inscribed disks of area $a_n$.
Since $\rho_n \rightarrow 0$ by assumption A1, all the arguments in Lemma 4.8 of \cite{gupta} go through.
Hence
\begin{align*}
\mbox{Prob} &\left\{\mbox{Number of nodes in cell $V$ $\geq$ $\frac{n a_n}{2}$, for every}\right.\\
&\left.\mbox{cell $V$ in the tessellation}\right\} > 1 - \delta_n\mbox{, }
\end{align*}
Also note that the
area of a disk of radius $\rho_n$ is lower bounded by $\frac{\pi {\rho_n}^2}{2}$, and hence the
result follows.
\end{proof}
\vspace{0.2cm}

Thus with high probability, each node can inject packets into the network at a rate no greater than $\frac{4}{\pi n {\rho_n}^2}$.
Thus,
\begin{equation*}
\lambda_n (i) \leq \frac{4}{\pi n {\rho_n}^2}
\end{equation*}
Substituting the above in \eqref{Lambdainter} results in the same $O\left(\frac{1}{n}\right)$ upper
bound on the per-node throughput. 
Thus we have the following generalized version of Proposition \ref{propmain1} that holds for any
$\rho_n$ that satisfies A1.

\begin{proposition}
\label{propmain2}
For any choice of $\rho_n$ that satisfies assumptions A1 to A4, 
the per-node throughput that can be achieved is $O(\frac{1}{n})$.
\end{proposition}
\endproof

Note that in obtaining the above result, we have assumed that each node transmits a packet just once.
If the packet is not received successfully at the receiver, no retransmission attempt is made. It is
easy to see that the conclusion in Proposition \ref{propmain2} holds even if a fixed number of
retransmissions are allowed, and this number does not scale with $n$.
Thus, under a realistic link layer model (assumption A2), if we do not scale the number of
retransmission attempts with $n$, and also keep the schedule length $K$ fixed, then the Gupta-Kumar result of per-node throughput of
$\Theta\left(\frac{1}{\sqrt{n\log{n}}}\right)$ does not hold.

\subsection{Per-node Capacity: Assumption A4 relaxed}
\label{seca1a4}
In the previous subsection, we saw that when straight line routing (proposed in \cite{gupta}) is used, the per-node throughput is
$O\left(\frac{1}{n}\right)$. 
In this sub-section, we show that this bound on the throughput also holds for a broad class of
routing schemes, i.e., Proposition \ref{propmain2} holds even when assumption A4 is relaxed.
In A4 we assume a straight line routing path between the source and
the destination nodes. However, we would like to know if a higher per-node
throughput can be achieved if we choose routing paths intelligently, i.e., not necessarily straight line 
paths between source-destination pairs. We only consider those routing paths in which
multi-hop communication takes place through packet-relaying between adjacent cells,
and there are no loops in the routing path.
The first requirement implies that communication is not possible between two nodes if they do not belong to adjacent cells,
while the second requirement implies that a packet does not visit a cell more than once on its path from source to destination.
Note that even with this restriction, we can account for a vast range of routing schemes.

Let us assume that a packet of connection $i$ takes a certain path (not necessarily straight line
path) from the source node to the destination node. Let ${\hat{L}}_{i}$ be the length of this path, i.e.,
the actual distance traveled by the packet along the surface of $S^2$. Then we have the following Lemma which is essentially a
generalization of Lemma \ref{lemmaHh}. In this Lemma, we show that at least a fixed fraction of hops for any routing scheme are over
a distance of $t\rho_n$ or longer for a fixed $t>0$. 

\begin{lemma}
\label{lemmartng}
Fix $W=40$, and $t=0.01$.
Assume {\em any} arbitrary routing path such that 
\begin{description}
\item[R1] Each hop is between nodes of two adjacent cells,
\item[R2] There are no loops in the routing path.
\end{description}
Then, under assumptions A1 to A3, R1 and R2, we cannot have $W$ consecutive hops of distance $t \rho_n$ or {\em less}. In other words, for any routing scheme
that satisfies A1 to A3, R1 and R2, at least one out of every $W$ consecutive hops is at least $t \rho_n$ long.
\end{lemma}
\begin{proof}
\begin{figure}
\epsfysize=4cm
\centerline{\epsfbox{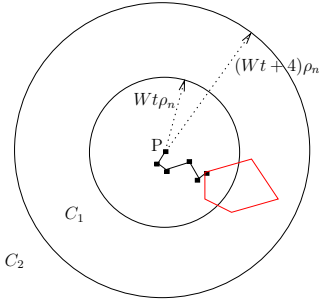}}
\caption{Bounding the number of hops of length $t \rho_n$ or less.}
\label{figrtng}
\end{figure}
By way of contradiction, assume that there exists some routing scheme that satisfies constraints A1 to A3, R1 and R2, and has $W$ consecutive hops, each of length $t \rho_n$ or less. 
Referring to Fig.~\ref{figrtng}, let point P be the initial location of the packet, and assume that the packet traverses $W$ hops from
this position. Since each of these hops is less than $t\rho_n$, the final location of the packet after traversing these $W$ hops is
inside a circle of radius $W t \rho_n$ centered at P. Let $C_1$ be this circle.
Clearly, every cell that the packet traversed during these $W$ hops must intersect circle $C_1$. Otherwise, the packet could not have
traversed that cell. We can easily find an upper bound on the number of cells that intersect $C_1$ as follows.
Consider another circle $C_2$ of radius $(W t+4)\rho_n$ centered at P. If a cell intersects $C_1$, it should lie {\em completely} inside
circle $C_2$. This is because the distance between any two points of a cell cannot exceed $4\rho_n$, and if a cell has at least one point
inside $C_1$, its other point cannot be more than $4\rho_n$ farther from the circumference of circle $C_1$.
The number of cells that lie completely inside $C_2$, denoted by $n_2$ is upper bounded as follows
\begin{align}
n_2 &\leq \frac{\pi \left((W t+4)\rho_n\right)^2}{\left(\frac{\pi {\rho_n}^2}{2}\right)} \nonumber \\
    &= 2 (W t+4)^2 \text{, }
\end{align}
where we have used Lemma \ref{lemmaEXP2} in the first step. For $W=40$ and $t=0.01$, we have $n_2 \leq 38.72$.  
Since 40 hops require at least 41 unique cells, and there are only up to 38 unique cells that the packet
can visit, we clearly have a contradiction.
Thus, there does not exist a routing scheme that satisfies A1 to A3, R1 and R2, and that has $W$ consecutive hops of $t \rho_n$ 
or less, where $W=40$, and $t=0.01$.
\end{proof}

Note that the arguments in the above Lemma are easier to understand in case of simple tessellations such as triangular, square, hexagonal, etc.
For example, consider a square tessellation (square-shaped cells), such that each side of the cell is $2\rho_n$ long. Thus, each cell contains a circle of
radius $\rho_n$, and is contained in a circle of radius $2\rho_n$. In this case, it is easy to see that circle $C_1$ defined in Lemma \ref{lemmartng}
cannot intersect more than 4 circles, and this happens in the neighborhood of one of the vertices of the tessellation. 
Consequently, for a square tessellation, we cannot have 4 consecutive hops, each of length $0.01 \rho_n$ or less.
In fact, the upper bound could be made tighter by noting that there cannot be more than 4 consecutive
hops that are strictly less than $0.5 \rho_n$.
Similarly, for a hexagonal tessellation such that each cell has sides of length $2\rho_n$,
we cannot have 3 consecutive 
hops, each of length strictly less than $0.5 \rho_n$. Thus the value of $W$ is much smaller, and the value of 
$t$ is much larger for simple tessellations. However when we consider general tessellations that
satisfy assumption A1, we note that the tessellation need not be as well-behaved as square or hexagonal. Yet, we
can obtain an upper bound on the maximum number of successive hops of arbitrarily small lengths. To do
this, we need to choose a fairly large (but finite and fixed) $W$, and a fairly small (but positive
and fixed) $t$.

In the next Lemma, we show that except for a small fraction of hops, all other hops have an interfering transmitter within a
distance of $(M+8)\rho_n$ of the receiver, and this fraction can be made arbitrarily small by choosing $M$ large enough.
\begin{lemma}
\label{lemmaMgen}
Fix $M>16$. Consider an arbitrary routing path between source and destination nodes such that R1 and
R2 are satisfied.
Let $N_i$ be the number of hops of connection $i$ such that
there is no simultaneous interfering transmission within a circle of radius $(M+8) \rho_n$ around
the receivers of those hops. To avoid tedious boundary conditions, let us not count the hops containing the source
and the destination nodes in $N_i$. Then,
\begin{equation*}
N_i \leq \frac{{\hat{L}}_i}{\rho_n} \left( \frac{2(1 + c_1)}{M}\right)\mbox{,}
\end{equation*}
where $\hat{L}_i$ is the length of the routing path and $c_1$ is the constant from Lemma 4.4 in \cite{gupta}.
\end{lemma}

\begin{proof}
\begin{figure}
\epsfxsize=8cm
\centerline{\epsfbox{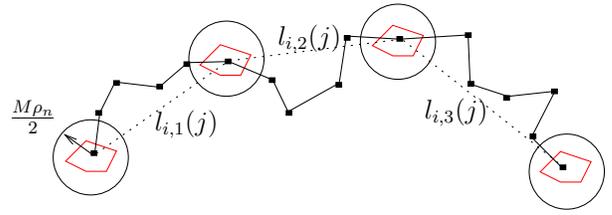}}
\caption{Arbitrary routing path.}
\label{figMgen}
\end{figure}
We use an approach identical to the one used to prove Lemma \ref{lemmaM}. 
Let $U_i^j$ be the set of all the cells of color $j$ that a packet of connection $i$ traverses.
Consider all the surrounding
circles of radius $M \rho_n$ around each of the cells in set $U_i^j$. 
For each cell, there are many such circles, since the only requirement is that
the center of the circle lie within the cell. A subset $V_i^j$ of $U_i^j$ is then formed as follows.
A cell from set $U_i^j$ is added to set $V_i^j$ if {\em none} of
its surrounding circles of radius $M \rho_n$ contains  any other color-$j$ cell either partially or fully.
Thus every cell in set $V_i^j$ is at least $M \rho_n$ distance away from any other color-$j$ cell.
Hence, any surrounding circle of radius $\frac{M \rho_n}{2}$ around a cell in set $V_i^j$ does not 
overlap with any surrounding circle of radius $\frac{M \rho_n}{2}$ around another cell in set $V_i^j$.
Let there be $N_{i}^{j}$ cells in set $V_i^j$,  and let $N_i$ be the sum of the number of cells in
$V_i^j$ for all $j$. We would like to determine an upper bound on $N_i$.

Referring to Fig.~\ref{figMgen}, consider the path taken by a packet of connection $i$ (solid line in
the figure). The path is piece-wise linear, with the locations of the relaying node of each hop as points of discontinuity.
Let $\hat{L}_i$ be the length of this routing path.
Let us rearrange the cells in set $V_i^j$ {\em in the
order in which a packet of connection $i$ traverses them}, 
and join them in this order (dotted lines in Fig.~\ref{figMgen}).
Let $l_{i,k}(j)$ be the length of the $k$th dotted line segment.
Define ${\hat{L}}_i^{'}(j)$ to be the total length of the dotted path. Then we have
\begin{equation}
{\hat{L}}_i^{'}(j) = \sum_{k} l_{i,k}(j) \leq {\hat{L}}_i
\label{lprime}
\end{equation}
\begin{figure}
\epsfxsize=8cm
\centerline{\epsfbox{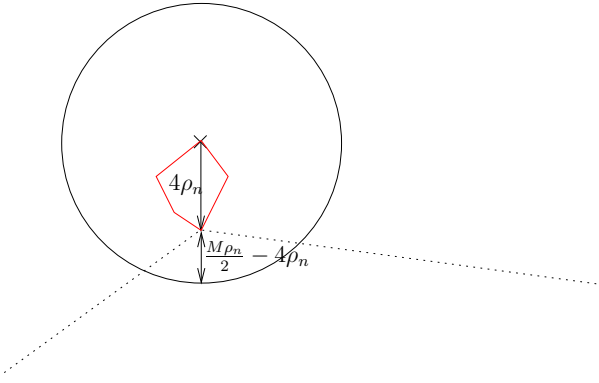}}
\caption{Bounding the number of hops $N_i$ of connection $i$ for which there is no interferer within
a distance $(M+8) \rho_n$ from the receiver with arbitrary (not straight line) routing.}
\label{figMgenlin}
\end{figure}
Consider a typical cell in set $V_i^j$ as shown in Fig.~\ref{figMgenlin}. 
As in the proof of Lemma \ref{lemmaM}, we first
determine the length of the minimum possible section of the routing path (in this case the dotted path) that lies inside the
surrounding circle. Since we know the total length of the dotted path, this gives us an upper bound on the number surrounding circles of
radius $\frac{M \rho_n}{2}$ that can be accommodated along the dotted path.

Referring to Fig.~\ref{figMgenlin}, the dotted path has its smallest section inside the surrounding circle of radius $\frac{M
\rho_n}{2}$
when the relaying node of the
cell, i.e., the point of discontinuity of the dotted path is located as close as possible to the circumference of the circle.
However, since this point belongs to the cell, it cannot be farther than $4\rho_n$ from any other point in the cell. But the center
of the surrounding circle must also lie inside the cell. Hence the point of discontinuity can be at the most $4\rho_n$ away from the
center of the circle. Hence the each portion of the
dotted path that lies within the circle can be no shorter than $\frac{M \rho_n}{2} - 4 \rho_n$. Thus, for each cell in $V_i^j$, at
least $(M-8)\rho_n$ portion of the dotted path must lie inside the surrounding circle. Let
$M-8>\frac{M}{2}$, i.e., $M>16$. Then the maximum number of such surrounding circles that can be accommodated along the dotted path,
i.e., the maximum number of cells in set $V_i^j$ is:
\begin{equation*}
N_{i}^{j} \frac{M \rho_n}{2} \leq \hat{L}_i^{'} \Rightarrow N_{i}^{j}  \leq \frac{2 \hat{L}_i^{'}}{M\rho_n}
\end{equation*}
Therefore for all the colors,
\begin{align*}
N_i &= \sum\limits_{j=1}^{1+c_1} N_{i}^{j} \\
&\leq\frac{\hat{L}_i^{'}}{\rho_n}\left( \frac{2(1+c_1)}{M} \right) \\
&\leq\frac{\hat{L}_i}{\rho_n}\left( \frac{2(1+c_1)}{M} \right)\text{using \eqref{lprime}.}
\end{align*}
Thus except for these $N_i$ hops, each of the hops of connection $i$ has at least one cell that has a
simultaneous on-going transmission
within a radius of $M \rho_n$ of the given cell. Or equivalently, within a radius of $(M+4+4)
\rho_n = (M+8)\rho_n$ of the {\em receivers}\, of these hops, there is at least one more
node that is transmitting simultaneously. This is because the maximum size of a cell is $4 \rho_n$.
And this proves the result.
\end{proof}

For any routing path, the number of hops is lower bounded by $\frac{\hat{L}_i}{8 \rho_n}$, since each hop can traverse a distance of
at the most $8 \rho_n$. Lemma \ref{lemmartng} shows that at least $\frac{1}{40}$th of these hops, i.e., at least
$\frac{\hat{L}_i}{320 \rho_n}$ hops of connection $i$ have transmitter-receiver pair separated by $0.01 \rho_n$ or more. 
As in Subsection \ref{seca1a5}, let $A_i$ be the set of hops of connection $i$ for which the received signal is at the most $P
(0.01 \rho_n)^{-\alpha}$. Then,
\begin{equation}
\label{newnAi}
|A_i| \geq \frac{\hat{L}_i}{320 \rho_n}
\end{equation}
Let $B_i$ be the set of hops of connection $i$ for which there is no simultaneous transmissions within a distance of 
$(M+8) \rho_n$ of the receiver.
Using Lemma \ref{lemmaMgen}, given $\epsilon >0$, 
we can choose $M$ large enough so that
\begin{align}
|B_i| &\leq \frac{\hat{L}_i}{\rho_n} \left( \frac{2(1+c_1)}{M}\right) \nonumber \\
      &\leq \frac{\hat{L}_i}{\rho_n} \epsilon
\label{newnBi}			
\end{align}
Then using \eqref{newnAi} and \eqref{newnBi},
\begin{align*}
|A_i \cap {B_i}^{c}| &\geq |A_i| - |B_i| \nonumber \\
& \geq \frac{\hat{L}_i}{\rho_n} \left( \frac{1}{320}\right) - \frac{\hat{L}_i}{\rho_n} \epsilon \nonumber \\
& = \frac{\hat{L}_i}{\rho_n} \left( \frac{1}{320} - \epsilon\right)
\end{align*}
If we pick $\epsilon =  1/640$, and choose $M=M_0$ corresponding to this choice of
$\epsilon$,
\begin{equation}
|A_i \cap {B_i}^{c}| \geq \frac{\hat{L}_i}{640 \rho_n}
\label{newnAB}
\end{equation}
Note that $A_i \cap {B_i}^{c}$ is the set of hops over which the received signal is no more than $P
(0.01 \rho_n)^{-\alpha}$, {\it and} there is at least one simultaneous transmission within a distance of $(M_0
+ 8) \rho_n$ of the receiver. This in turn means that for these hops, the SINR is upper bounded by
\begin{align}
\label{newsinr}
SINR &\leq \frac{P (0.01 \rho_n)^{-\alpha}}{P \left((M_0 + 8)\rho_n\right)^{-\alpha}} \\
\label{newsinr}
&= 100^\alpha \left( {M_0 + 8} \right)^{\alpha} \nonumber \\
&= \beta_1 \nonumber
\end{align}
The above upper bound on SINR may seem large due to $100^\alpha$ term. However, this term
appears because we picked up a conservative hop length of $0.01 \rho_n$ in proving Lemma
\ref{lemmartng} to account for all possible tessellations. For simple tessellations such as square and
hexagonal, we can use a minimum hop length of $\rho_n$, and obtain a tighter bound on
SINR. However, as far as the asymptotic nature of the solution is
concerned, the values of these constants do not matter.
Thus we have proved the following Proposition which is a generalization of Proposition \ref{propphi} when A4 is relaxed.
\begin{proposition}
\label{propphinew}
There exists a fixed constant $M_0$, that does not depend on $n$, such that for at least
$\hat{L}_i/640\rho_n$ hops of connection $i$, the SINR is less than a fixed constant $\beta_1$ given by
\begin{equation*}
\beta_1    = 100^\alpha \left( {M_0 + 8} \right)^{\alpha}
\end{equation*}
As per assumption A2, since the SINR is upper bounded by a fixed constant $\beta_1$, the
probability of successful packet reception is also upper bounded by a fixed
constant $\phi(\beta_1) < 1$.
\end{proposition}
\endproof

We can now use the above Proposition, and proceed as in Subsection \ref{seca1a5} to determine the end-to-end throughput of a
connection. For this, we note that at least $\hat{L}_i/640\rho_n$ hops of connection $i$ have an SINR that is upper bounded by a
constant $\beta_1$, i.e., a packet success probability that is upper bounded by a constant $\phi(\beta_1)$. 
Hence, replacing $L_i$ by ${\hat{L}}_i$ in \eqref{Lambdafirst} to \eqref{Lambda} we obtain the throughput a connection, $\Lambda_n$ as
\begin{equation*}
\Lambda_n          \leq \lambda_n (i)\, {\bf E}_L \left[ \delta^{{\hat{L}}_i} \right]
\end{equation*}
Recall that $L_i$ denotes the length of the straight line joining the source-destination node pair of connection $i$,
and hence ${\hat{L}}_i \geq L_i$. Also, since $\delta \leq 1$, we get
\begin{equation*}
\Lambda_n \leq \lambda_n (i)\, {\bf E}_L \left[ \delta^{L_i} \right]
\end{equation*}
After this stage, we use \eqref{deltaL}, and proceed along the same
lines as Subsection \ref{seca1a5}, and obtain the same asymptotic upper bound of $O\left(\frac{1}{n}\right)$
as in Propositions \ref{propmain1} and \ref{propmain2}. 
\begin{proposition}
\label{propmain2new}
For any choice of $\rho_n$ that satisfies assumptions A1 to A3, and for any routing scheme that
satisfies R1 and R2, the per-node throughput that can be achieved is $O(\frac{1}{n})$.
\end{proposition}
\endproof
Thus the asymptotic per-node throughput is
$O\left(\frac{1}{n}\right)$ even with arbitrary routing, i.e., when assumption A4 is relaxed.

\subsection{{Per-node Capacity: Assumption A3 relaxed}}
In the previous subsection, we saw that even when assumption A4 was relaxed, 
for a broad range of routing schemes the achievable per-node throughput is 
$O\left(\frac{1}{n}\right)$. To prove this, we showed that at least a fixed fraction of hops have the transmitter-receiver separation
of no more than $t \rho_n$ for some fixed $t>0$, {\em and} there exists an interfering transmitter within a distance of $(M+8) \rho_n$ 
of receivers of these hops for some fixed, and finite $M$. This enabled us to upper bound the SINR (and hence the packet success probability) of
these hops, and this in turn led to the $O\left(\frac{1}{n}\right)$ per-node throughput. In this sub-section, we show that if we
reduce the extent of spatial reuse via scheduling, then we can get better per-node throughput. Reduced spatial reuse results in lower
interference from other ongoing transmissions, and this in turn improves SINR.

Recall that in proving Lemma \ref{lemmaMgen}, we showed that there exists an interfering 
transmitter within a distance of $(M+8) \rho_n$ of at least a certain fraction of hops of connection $i$, and this fraction can be made as
close to one as desired by choosing $M$ appropriately large. To prove this, we considered one color at a time, and determined the
maximum number of cells of that color that belong to connection $i$ such that these cells {\em do not} have a simultaneous transmission within a radius $M \rho_n$ of themselves. We then
summed up over all the colors to determine the maximum number of cells of connection $i$ with this property. Due to assumption A3,
the number of colors used during scheduling, i.e., the length of the schedule is $1+c_1$, and $c_1$ is a constant. 

To relax assumption A3, let us assume that the
total number of colors that are used during scheduling is $K_n$, i.e., the schedule length is not constant, but it scales with $n$.
Also assume that $K_n$ goes to infinity as $n$ scales goes to infinity as $n$ scales. This implies that as the number of nodes in the network increases, the schedule become 
progressively longer, or equivalently, the extent of spatial reuse decreases with increasing network size. Due to this choice of
$K_n$, i.e., due to reduced spatial reuse, we cannot
use Lemma \ref{lemmaMgen} to upper bound $N_i$. One way to bound $N_i$ would be to consider surrounding
circles of radius $M K_n  \rho_n$ instead of $M \rho_n$, so that $N_i$ can be made arbitrarily small by choosing $M$ large as before.
However, the problem with this approach is that it guarantees that the fraction of hops of connection $i$ with an interfering
transmitter within a distance of $(M K_n + 8)\rho_n$ can be made as close to one as desired by choosing $M$ large. However, our
ultimate goal is to upper bound the SINR, and this was done in two parts; first by showing that at least a certain fraction of hops
have transmitter-receiver separated by $f_n$, and that for except for a small fraction of hops, all the hops have an interferer
within a distance of $g_n$. By showing that $f_n = \Theta(g_n)$, we were able to show that the SINR is upper bounded by a constant.
Hence in addition to showing that there is an interfering node within $(M K_n + 8)\rho_n$ of the receiver,
we must also show that at least a fixed fraction of hops have transmitter and receiver separated by no more than $t K_n \rho_n$ for
some fixed positive $t$. Then we can claim an upper bound on the SINR of these hops. However there is no analog of Lemma
\ref{lemmartng} in this case. The reason being that we cannot claim that there are at the most
$W$ consecutive hops such that each of them is shorter than $t K_n \rho_n$ because $K_n \rightarrow \infty$.
Consequently, we can no longer claim that the 
per-node capacity of the network is $O\left(\frac{1}{n}\right)$.

Thus by choosing a more conservative scheduling policy, i.e., by choosing a schedule of length $K_n$ such that $K_n$ goes to infinity
as $n$ scales, we reduce the interference on the links and improve the link SINR. However note that a reduced spatial reuse has an
impact on the per-node throughput.
We know from Theorem 4.1 in \cite{gupta}, that the straight line routing scheme achieves a per-node
throughput of $\frac{c}{n \rho_n (1+c_1)}$, where $c$ is a constant. In \cite{gupta}, this throughput was shown to be
achievable by determining the number of connections that pass though a given cell, and using the fact
that each cell gets a transmission opportunity once every $1+c_1$ slots, i.e., the schedule length is $1+c_1$.
For a general schedule length $K_n$, the throughput is given by
$\frac{c}{n \rho_n K_n}$. To ensure connectivity, $\rho_n$ cannot decay any faster than $\sqrt{\frac{\log{n}}{n}}$, and hence the per
node throughput is upper bounded by $\frac{c}{K_n \sqrt{n\log{n}}}$.
Thus, to overcome the cumulative packet error probability over an asymptotically long connection, we must compromise on the spatial
reuse in the network. By employing a schedule of asymptotically long length, we can reduce the extent of interference received by a link
from other ongoing transmissions. This improves the link SINR. The effect of reduced spatial reuse is that now the per-node
throughput is $O\left(\frac{1}{K_n \sqrt{n \log{n}}}\right)$ instead of $O\left(\frac{1}{n}\right)$.
Note that for this to be true, $K_n$ has to be $\omega(1)$.

We have not shown that the above throughput can be achieved, i.e., 
with the above choice of $K_n$, and under assumptions A1 and A2, either the straight-line routing
scheme in \cite{gupta} or any other
routing scheme {\em achieves} a throughput of $\Theta\left(\frac{1}{K_n \sqrt{n \log{n}}}\right)$. This is an
interesting problem, and will be addressed in a separate work.

\subsection{A Remark about Retransmission Strategy}
In our analysis so far, we have assumed that over each hop, a single attempt is made to transmit the packet.
If the transmission fails, then the node does not make a retransmission attempt. In reality, a multi-hop
link layer employs a hop-by-hop acknowledgment and retransmission strategy in which the transmitter retransmits the packet if its
previous transmission fails. If we assume that an infinite number of retransmissions are allowed for
each packet, then the packet will eventually be successfully relayed over each hop. However, most practical link
layer protocols have a fixed upper bound on the number of allowable retransmissions to avoid unbounded
delays. 
If no more than $R$ retransmissions are allowed over each hop, and if $p$ is the probability of packet
success over each transmission, then the probability that the packet will be transmitted successfully
within $R$ attempts is simply given by $1 - {(1-p)^{R}}$. This quantity is strictly less than one for
finite $R$. Thus, there is an upper bound on the packet success probability which is strictly less than
one. Note that for $R=1$, this probability reduces to $p$, which is the case we have studied. With a
constant $R$ that does not scale with $n$, all our previous arguments still go through
with slight modifications. We can now think of replacing $p$ as the packet success probability by 
$1 - {(1-p)^{R}}$, which is still a constant independent of $n$, and strictly less than one.
As a result, the per-node throughput still remains $O\left(\frac{1}{n}\right)$.

Note that {\it asymptotically}, allowing infinite number of retransmissions at the link level is equivalent
to assuming that the packet will be successfully received in a single transmission if the SINR is
greater than $\beta$. 
However a realistic link layer protocol allows only a fixed bounded number of retransmission attempts.
Besides, using a more conservative scheduling policy, we could also scale the number of allowable
retransmissions so that the probability of transmission success over each link approaches one
asymptotically. 
The impact of a retransmission scheme that scales with $n$, on the network capacity is
an interesting problem that will be addressed in an extension of this work.

\section{Conclusion and Extensions}
\label{future}
The key observation in this paper is that
for a broad range of routing and scheduling schemes (including the one proposed in \cite{gupta}),
the number of hops of each connection scales to infinity with $n$.
In a realistic link layer model, for a given modulation and coding scheme, the
probability of packet loss over any given link is non-zero for finite SINR values.
With the above link layer model, we show that for a broad class of routing and scheduling policies
we cannot achieve a per-node throughput of more than $O \left( \frac{1}{n}\right)$ due to the
cumulative probability of packet loss over all the hops of the connection. 
However, it is possible to improve the link SINR by using a more conservative scheduling policy with a
lower spatial reuse. In this case the per-node throughput is $O\left(\frac{1}{K_n \sqrt{n \log{n}}}\right)$,
where $K_n$ is the length of the schedule, i.e., each cell gets a transmission opportunity once every
$K_n$ slots, and $K_n$ goes to infinity as $n$ scales. Thus the capacity of random ad-hoc networks
scales even more pessimistically than what was previously thought.

Although we have shown that with such a choice of $K_n$, the per-node throughput is
$O\left(\frac{1}{K_n \sqrt{n \log{n}}}\right)$,
whether this throughput can indeed be {\em achieved} by a routing and scheduling scheme needs to be studied, and
will be addressed in an extension of this work.
We would also like to study how the throughput scales when we allow the number of retransmissions to
scale with $n$.
The work in \cite{gupta} assumes that nodes pick source destination pairs at random, and then use
multi-hop communication. The authors then note that each cell has to relay packets of other
connections, and it is this relaying load that determines how fast a node can inject its own packets
into the network. Thus the results are strongly dependent on the any-to-any communication paradigm,
i.e., a network architecture in which any node may wish to communicate with any other node. On the
other hand, our results are mainly based on the
impact of interference and cumulative packet error probability on the network throughput. 
We note that the scheduling
strategy and the number of hops are the key parameters that may determine the network throughput,
and these parameters {\em do not} require any additional assumptions on the communication paradigm
(any-to-any, many-to-one, arbitrary). As the network size increases, paths are likely to have an
asymptotically large number of hops with high probability.
Thus our results can be generalized to more generic network architectures
such as \cite{franc,dousse}, and this will also be addressed in an extension of this work.

\section*{Acknowledgments}
This work was supported in part by the Indiana Twenty First Century Fund through the Indiana Center
for Wireless Communication and Networking, and by the National Science Foundation Grant No. 0087266.
We would like to thank Prof. Ravi Mazumdar (University of Waterloo, Canada), and Prof. Ness Shroff 
(Purdue University, USA) for
their valuable suggestions that greatly helped us improve the quality of this manuscript.

\section*{Appendix}
\begin{lemma}
\label{lemmaEXP}
If we pick two points randomly and uniformly on the surface of sphere $S^2$, and connect them by a
line (drawn along the surface of $S^2$), and if ${\bf L}$ is the random variable corresponding to the
length of this line, then for $0 < \delta < 1$,
\begin{equation*}
{\bf E} \left[ \delta^{\bf L} \right] = \frac{2\pi\left(1 +
\delta^{\frac{\sqrt{\pi}}{2}}\right)}{4\pi + (\log \delta)^2}
\end{equation*}
\end{lemma}
\vspace{0.2cm}

\begin{proof}
Since both the points are picked randomly and uniformly on the surface of $S^2$, without loss of
generality, assume that one point is located at the north pole of the sphere. Then, we can find the
probability that the other point is located within a distance of $l$ from the north
pole. This is the same as the event $A= \left\{{\bf L} \leq l \right\}$, and its probability 
equals the ratio of the
area of the shaded region in Fig.~\ref{figEXP} to the area of the sphere, which is one.
Note that all the distances are measured along the surface of the sphere. 
\begin{figure}
\epsfysize=4cm
\centerline{\epsfbox{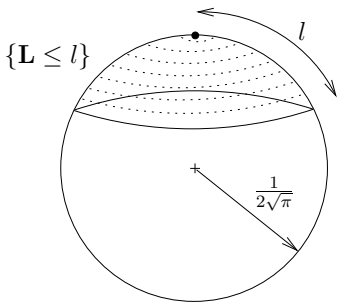}}
\caption{Probability distribution of ${\bf L}$.}
\label{figEXP}
\end{figure}
We have,
\begin{equation}
\mbox{Area}\left\{\mbox{Shaded region}\right\}
= \frac{1}{2} \left( 1 - \cos{\left(2\sqrt{\pi}l\right)}\right)
\label{area}
\end{equation}
Therefore the distribution function of ${\bf L}$, i.e., $F_{L} (l)$ is
\begin{equation*}
F_{L} (l) = \frac{1}{2}(1 - \cos{\left(2\sqrt{\pi}l\right)}) \mbox{.}
\end{equation*}
Note that $0 \leq {\bf L} \leq \sqrt{\pi}/2$. 
The density function of ${\bf L}$ is given by,
\begin{equation*}
f_{L} (l) = \sqrt{\pi}\sin{\left(2\sqrt{\pi}l\right)} \mbox{. }
\end{equation*}
Now we can compute ${\bf E}[{\delta}^{\bf L}]$ as follows.
\begin{align}
{\bf E}\left[{\delta}^{\bf L}\right] 
&= \int\limits_{0}^{\frac{\sqrt{\pi}}{2}}  {\delta}^{l} \sqrt{\pi}\sin{\left(2\sqrt{\pi}l\right)} dl \nonumber \\
&= \frac{1}{2}\int\limits_{0}^{\pi}  {\gamma}^{x} \sin{x} dx\mbox{, } \quad \mbox{ where $\gamma =
{\delta}^{\frac{1}{2\sqrt{\pi}}}$} \nonumber \\
&= \frac{1}{2} I
\label{EdeltaL}
\end{align}
Solving for $I$ using integration by parts,
\begin{equation*}
I=\frac{1 + {\gamma}^{\pi}}{1 + (\log{\gamma})^2} 
\end{equation*}
Replacing $\gamma$ by ${\delta}^{\frac{1}{2\sqrt{\pi}}}$, and substituting the above in \eqref{EdeltaL},
\begin{equation}
{\bf E}\left[{\delta}^{\bf L}\right]  = \frac{2\pi\left(1 + {\delta}^{\frac{\sqrt{\pi}}{2}}\right)}{4\pi + (\log{\delta})^2} 
\end{equation}
\end{proof}
\vspace{0.2cm}
Using \eqref{area} from the above Lemma, we also have the following lemma
for the area of a disk of radius $\rho_n$ on $S^2$.
\begin{lemma}
\label{lemmaEXP2}
If $\rho_n$ is the radius of a disk (measured along the surface of $S^2$), and the area of this disk is
$a_n=100\log{n}/n$, then
\begin{equation}
\frac{\pi {\rho_n}^2}{2} \leq a_n \leq {\pi}{\rho_n}^2 \mbox{.}
\label{area2}
\end{equation}
\end{lemma}
\vspace{0.2cm}

\begin{proof}
Using \eqref{area} with $l=\rho_n$ gives
\begin{align*}
a_n &= \frac{1}{2}\left(1-\cos{\left(2\sqrt{\pi}\rho_n\right)}\right) \\
    &= \sin^2{\left(\sqrt{\pi}\rho_n\right)}
\end{align*}
For $\theta \leq {\pi}/4$,
\begin{equation*}
\frac{\theta}{\sqrt{2}} \leq \sin{\theta} \leq \theta
\end{equation*}
From this, \eqref{area2} follows if $n$ is large enough so that $\rho_n \leq \sqrt{\pi}/4$.
The latter can always be ensured since
\begin{equation*}
a_n = \frac{100 \log{n}}{n} = \sin{\left(\sqrt{\pi}\rho_n\right)}
\end{equation*}
\end{proof}
\vspace{0.2cm}


\end{document}